\documentstyle[aps,aps10,psfig,prl,multicol]{revtex}
\def\be{\begin{equation}}
\def\ee{\end{equation}}
\def\bea{\begin{eqnarray}}
\def\eea{\end{eqnarray}}

\begin{document}
\title{Experimental Evidence of Time Delay Induced Death in Coupled
Limit Cycle Oscillators}
\author{D.~V.~Ramana Reddy, A. Sen, and
G.~L.~Johnston\cite{GLJadd}}
\address{Institute for Plasma Research, Bhat, Gandhinagar 382428, India}
\maketitle

\begin{abstract}
Experimental observations of time delay induced amplitude death in
a pair of coupled nonlinear electronic circuits that are
individually capable of exhibiting limit cycle oscillations are
described. In particular, the existence of multiply connected
{\it death islands} in the parameter space of the coupling
strength and the time delay parameter for coupled identical
oscillators is established. The existence of such regions was
predicted earlier on theoretical grounds in [Phys. Rev.  Lett.
{\bf 80}, 5109 (1998); Physica {\bf 129D}, 15 (1999)]. The
experiments also reveal the occurrence of multiple frequency
states, frequency suppression of oscillations with increased time
delay and the onset of both
in-phase and anti-phase collective oscillations.\\


\end{abstract}

\begin{multicols}{2}
Coupled limit cycle oscillator models have been extensively
studied in recent years because of the useful insights they
provide into the collective behaviour of many physical, chemical
and biological systems \cite{KN:87,HGEK:97,CE:89,CS:93a}. One of
the simplest such models, the so-called Kuramoto model
\cite{KN:87}, which retains only the phase information of each
oscillator and is valid in the limit of weak mutual couplings,
displays a spontaneous transition to a synchronized collective
state of a single frequency when the coupling strength exceeds a
critical value. Similar collective behaviour is observed
in many natural systems, such as the synchronous flashing of
fireflies, the phase locking of cardiac pacemaker
cells, and the collective chirping of crickets \cite{CS:93a}.
Phase locking has also been demonstrated experimentally in arrays
of coupled nonlinear electronic circuits \cite{BL:96}. When the
coupling becomes stronger, amplitude effects become important and
give rise to other interesting collective states, such as that of
{\it amplitude death} in which the various oscillators pull each
other off their periodic states and collapse to a state of zero
amplitude. The condition for such a state to occur is for the
oscillators to have a broad dispersion in their
natural frequencies and for the coupling strength to exceed a
threshold value. Thus, as has been pointed out in a number of
theoretical studies \cite{AEK:90}, a collection of identical
limit cycle oscillators cannot display {\it amplitude death}.
However, more recent investigations \cite{RSJ:98} indicate that the
presence of finite propagation time delays in the coupling
removes this restriction and predict the possibility of inducing
the death state even in a system of two coupled identical limit
cycle oscillators. Time delay is ubiquitous in most physical
systems due to finite propagation speeds of signals, finite
chemical reaction times, finite response times of synapses, etc.,
and its influence on the collective dynamics of coupled systems
can have wide-ranging implications \cite{Strog:98}. It is
important therefore to establish the
experimental feasibility of such a death phenomenon.

In this Letter we present experimental observations on time delay
induced death in two coupled nonlinear circuits that are
individually capable of exhibiting limit cycle oscillations. A
specially designed digital delay line gives precise control over
the delay time in the coupling and permits us to explore a large
area of parameter space. Observations on the phenomenon of {\it
amplitude death} have recently been reported \cite{HFRPO:00} for
a pair of optothermal oscillators that are thermally coupled and
for which the occurrence of death for strong couplings and its
relation to Hopf bifurcations of the uncoupled and coupled
systems have been experimentally verified. However, as the
authors themselves point out, their experimental results are not
conclusive about the role of delay in the death phenomenon and in
particular they have not investigated the phenomenon of death
islands or, indeed, that of the multiplicity of death islands
predicted by the theory of time delay induced death
\cite{RSJ:98}. In our experiments we provide clear evidence of time delay
induced death islands and their multiple connectedness in the
parameter space defined by the coupling strength, time delay and
frequency, for the case of two coupled identical oscillators. We
also find the existence of multiple frequency states, frequency
suppression of oscillations with increased time delay and the
onset of both in-phase and anti-phase collective
oscillations. 

\begin{figure}
\narrowtext
\centerline{\psfig{file=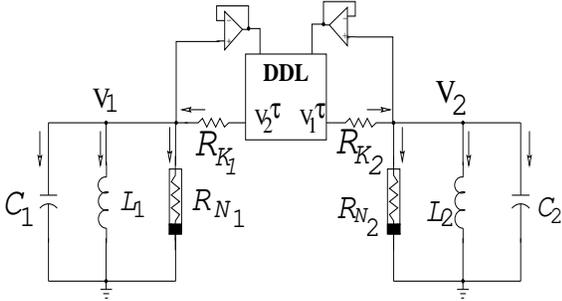,width=7.5cm,height=4cm}}
\caption{The circuit diagram of two delay coupled limit cycle
oscillators consisting of a $C$, $L$ and a nonlinear resistance
$R_N$ connected in parallel. The OP-AMPs are buffer amplifiers
and ideally draw no input currents. The digital delay line
outputs $V_{1,2}^{\tau} = V_{1,2}(t-\tau)$.} 
\label{FIG:ckt}
\end{figure}

\begin{figure}
\centerline{\psfig{file=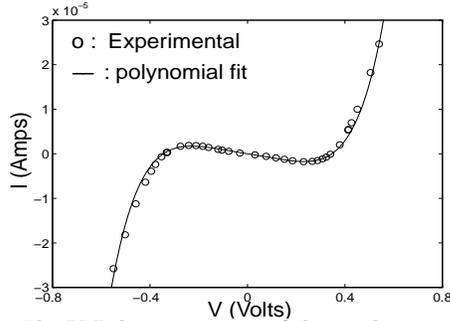,width=6cm,height=4.5cm}}
\caption{The V-I characteristics of the nonlinear component
$R_{N}$. The continuous line is a polynomial fit of the
experimental points.} \label{FIG:gv}
\end{figure}

\begin{figure}
\centerline{\psfig{file=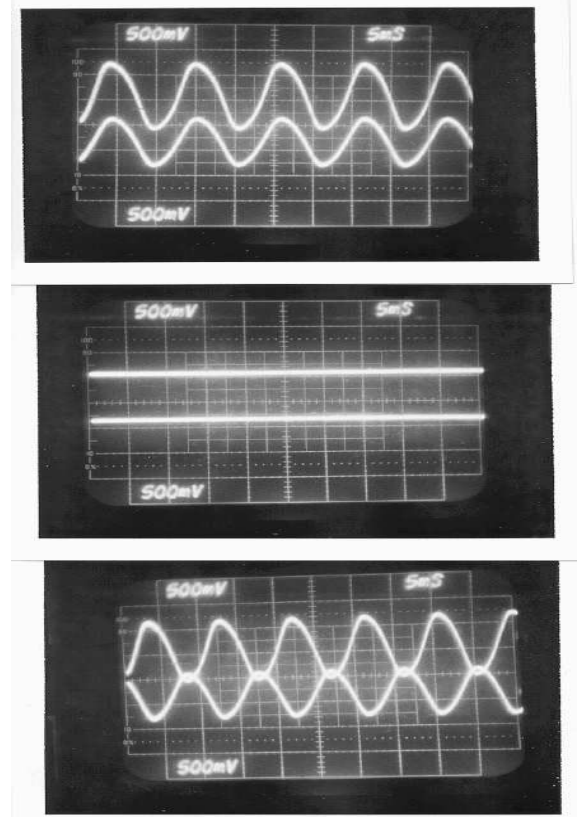,width=7.5cm,height=11cm}}
\caption{Oscilloscope traces of the temporal
behavior of the voltages of the two delay coupled oscillators for
different delay times with $K = 1000 s^{-1}$ and  $\omega = 837
s^{-1}$. Time is marked in units of $5$ ms along the horizontal
axis. The oscillators show (top panel) in-phase locking for $\tau
= 0.514 $ ms, (middle panel) amplitude death for $\tau = 2.0$
ms, and (bottom panel) anti-phase locking for $\tau = 4.428$ ms.}
\label{FIG:5.v1inanti}
\end{figure}

\begin{figure}
\centerline{\psfig{file=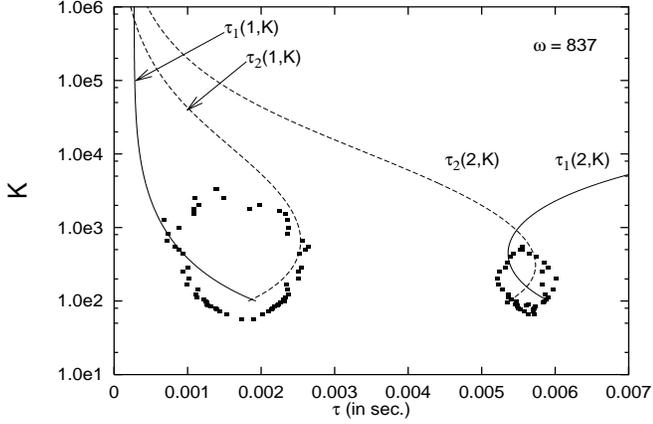,width=8cm,height=6cm}}
\caption{Multiply connected death islands in \(K-\tau\) space for
$\omega = 837 s^{-1}$. Solid squares mark the experimental points
while the solid and dashed lines are theoretical curves obtained
from model equation (1).} \label{FIG:isle}
\end{figure}

Our experimental system, schematically shown in
Fig.~\ref{FIG:ckt}, consists of two nonlinear $LCR$ circuits
coupled through a digital delay line (DDL). The individual oscillator
circuits are a variant of the so-called Chua circuit, which has
been widely employed in a number of nonlinear dynamical studies
\cite{Chua:93}. The nonlinear resistive element $R_{N}$ has the
typical \(V-I\) characteristic, with negative resistance around
the origin and positive slopes away from the origin as shown in
Fig.~\ref{FIG:gv}. This enables each individual oscillator
circuit in the uncoupled state to sustain limit cycle
oscillations with a characteristic frequency of $1/\sqrt{LC}$.
The two channel DDL taps the input signal at
a certain rate using an analog to digital converter (in this case
an ADC0809 with a conversion rate of $100 \mu $s) and stores it
in a random access memory (RAM) bank. The stored bits are read
using a logical circuit and converted back to an analog signal to
be fed back into the circuit. The coupling between the
oscillators is linear, resistive and proportional to the
difference in the signal strengths of the two oscillators with a
time delay. The coupling strength is varied by changing the
resistances ($R_K$) that join the two oscillators. By using the
current summation rules at nodes $V_{1}$ and $V_{2}$, a
theoretical description of the circuit can be given as, 
\be
\label{EQN:model1} \ddot{V}_{i} + g(V_i)~\dot{V}_{i} +
\omega_{i}^2 ~V_{i} = K_{i} [\dot{V}_{j}(t-\tau) -
\dot{V}_{i}(t)], 
\ee
where $i,j=1,2, i \ne j$ and $g(V_i) =(1 /
C_{i}) ( - 0.95 + 0.66 V_i^2 + 322.0 V_i^4 ) \times 10^{-5}$
Amp/F, the nonlinear damping factor, is obtained from an
empirical fit of the \(V-I\) characteristics of the nonlinear
resistive element as shown in Fig.~\ref{FIG:gv}.
The frequencies
are $\omega_j = 1 /\sqrt{L_j C_j}$ and the coupling strengths
$K_j = 1 / (C_j R_{K_j})$. For our experiments on identical
oscillators we have fixed $C_{1}= C_{2} = 0.1 \mu F$ and varied
$L = L_{1} = L_{2}$ and $R = R_{K_1} = R_{K_2}$ such that 
$\omega = \omega_{1} =\omega_{2}$ spans the 
range of $100 s^{-1}$ to $1000 s^{-1}$. 
%
%
%
The coupling strength varies in the range of $10 s^{-1}$ to $10^6 s^{-1}$.
Examples of our experimental
results for the time evolution of the oscillator voltages as a
function of the delay parameter are shown in
Fig.~\ref{FIG:5.v1inanti} for $K= 1000 s^{-1}$ and $\omega = 837
s^{-1}$. The identical oscillators acquire an in-phase locked
state (for $\tau = 0.514$ ms), death (for $\tau = 2$ ms) and an
anti-phase locked state (for $\tau = 4.428$ ms). A series of such
observations are recorded for a fixed frequency and at different
fixed values of $K$ by varying the time delay parameter. The
output voltage $V$ of the oscillator is monitored on the
oscilloscope up to the point where it shows a sudden drop in
value by several orders of magnitude, i.e., nearly to zero. The
value of $\tau$ is further increased up to the point where the
signal from the oscillator again suddenly revives to a large
value. An entire parameter space in \(K - \tau \) for amplitude
death is thereby explored.
In Fig.~\ref{FIG:isle} the solid squares are the experimental data
points marking the loci of the death regions. They clearly reveal
closed {\it death island} regions in the plane of the coupling
strength, $K$, and the time delay, $\tau$, for the common
frequency of $\omega = 837 s^{-1}$. 
%
%
%
%
The solid and dashed lines
are analytic curves obtained from a linear stability analysis of
the origin for the model equation (\ref{EQN:model1}).
These curves are derived in a standard manner \cite{RSJ:98}
from the characteristic equation for the eigenvalue $\lambda$, namely,
$\lambda^2 (\lambda-\frac{a_1}{C} + K)^2 + 2 \omega^2
(\lambda-\frac{a_1}{C} + K) \lambda + \omega^4 - K^2 \lambda^2
e^{-2\lambda \tau} = 0$, and are given by,
\bea
\label{EQN:DI1}
\tau_1(n,K)
&=& \frac{(n-1)\pi+\cos^{-1}(1-\frac{a_1}{C K})}
                    {-\frac{A}{2} + \sqrt{(\frac{A}{2})^2+\omega^2}}, \\
\label{EQN:DI2}
\tau_2(n,K) &=&
\frac{n\pi-\cos^{-1}(1-\frac{a_1}{C K})}
                    {\frac{A}{2} + \sqrt{(\frac{A}{2})^2+\omega^2}},
\eea
where $A = \sqrt{\frac{2 a_1}{C} K - (\frac{a_1}{C})^2}$,
$a_1=0.95 \times 10^{-5}$ and $n = 1, 2, \ldots$. The region
enclosed by the intersection of $\tau_1(1,K)$ and $\tau_2(1,K)$
is the primary death island and that enclosed between
$\tau_1(2,K)$ and $\tau_2(2,K)$ is the secondary (higher order)
death island. We observe that the experimental observations not
only display topological similarity to the analytic predictions
but are also in good quantitative agreement with them. The
deviations observed are primarily due to two reasons. First, the
analytic domains are sensitive to the values of the coefficients
used in fitting the \(V-I\) characteristic of the nonlinear
resistive element. The relative downward shift and larger width
of the experimental islands compared to the analytic region are
due to these estimation errors.
Second, the deviation at very high
values of $K$ (and hence low $R_{K}$) is related to limitations
on the maximum current that can flow through $R_{K_j}$ for the
rated biased  voltages of our experimental circuit. The existence
of the higher order island is particularly significant since it
confirms the theoretical finding of multiple connectivity of the
stability region of the origin \cite{RSJ:98}. 
%
\begin{figure}
\centerline{\psfig{file=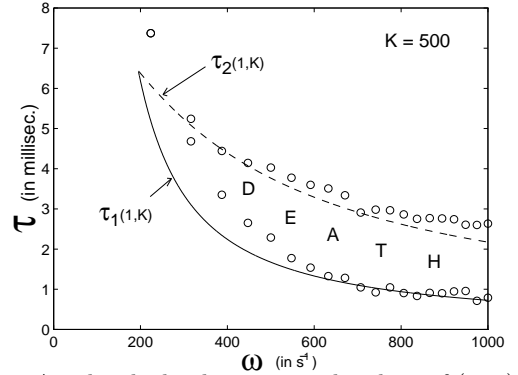,width=7cm,height=5cm}}
\caption{Amplitude death region in the plane of $(\omega,\tau)$
for a fixed coupling strength of $K = 500 s^{-1}$.}
\label{FIG:ftau}
\end{figure}
In Fig.~\ref{FIG:ftau}, the experimental death region is shown at a
constant coupling strength $K = 500 s^{-1}$ for different values
of the frequency. The death region ceases to exist below a
certain threshold frequency, in accordance with theoretical
predictions \cite{RSJ:98}.
On either side of this region, there
are phase locked states exhibiting in-phase locked oscillations
for small time delays and anti-phase locked oscillations for
longer time delays.

\begin{figure}
\centerline{\psfig{file=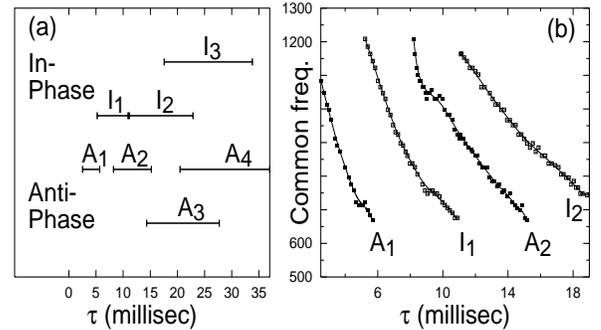,width=8cm,height=4.5cm,angle=270}}
\caption{(a) Coexistence of in-phase and anti-phase locked
states, and (b) suppression of the phase locked states as $\tau$
is increased for $K= 1000 s^{-1}$ and $\omega = 837 s^{-1}$.}
\label{FIG:coexsup}
\end{figure}

Another important characteristic of time delay systems in general
\cite{NSK:91,KPR:97}, and delay coupled limit cycle oscillators
in particular \cite{RSJ:98}, is the existence of multiple phase
locked states. When the oscillators are identical, these phase
locked states coincide with in-phase (phase difference, $\phi =
0$) and anti-phase ($\phi = \pi$) solutions. In
Fig.~\ref{FIG:coexsup}(a), the domains of existence of these
solutions in $\tau$ space are indicated. 
Here $I_n$ represent the in-phase solutions and $A_n$ represent
the anti-phase solutions. These domains have been determined
experimentally. For a fixed $\tau$ the various existing multiple
states can be accessed with slight perturbations to the system,
i.e., by slightly changing the initial conditions. We have also
observed hysteretic behaviour between the various branches.
In Fig.~\ref{FIG:coexsup}(b), the frequencies of the first few
multiple states are plotted for increasing values of $\tau$. Note
that the collective states can be much 
higher than the individual oscillator frequencies and that 
%
%
\noindent
their magnitude decreases with
increasing values of time delay. Such a phenomenon of time delay
induced frequency suppression has been discussed earlier in the
context of phase only oscillators \cite{NSK:91}.\\

In conclusion, we have experimentally demonstrated for the first
time the existence of delay induced amplitude death in a pair of
coupled identical nonlinear oscillators \cite{RSJ:98}. We have
also demonstrated the coexistence of in-phase and anti-phase
states with multiple frequencies and the suppression of the common
frequency with time delay. The observation of multiply connected
death islands (for identical oscillators) in the parameter space
of the coupling strength and time delay is our most important
result. This result not only bears topological
resemblance to earlier model calculations \cite{RSJ:98}, 
but also shows good quantitative agreement with analytic estimates
obtained from a theoretical model of the experimental circuits.
It should be noted that our experimental system of coupled
nonlinear electronic circuits (and its theoretical model) is
capable of richer nonlinear behaviour and goes beyond the simple
normal form model used in previous theoretical studies
\cite{RSJ:98}. In this sense the results on delay induced 
amplitude death can be
said to have a broader validity and therefore wider applicability
to physically realistic systems.\\

We now briefly discuss a few interesting directions for future
experiments that are relevant to our present results as
well as to practical applications. A natural next step is to
couple more than two oscillators to explore some of the other
theoretical predictions discussed in \cite{RSJ:98}. The principal
technical challenge and the major expensive elements for such an
endeavor are the construction of additional delay lines with
enhanced number of channels. Such an effort is presently
underway.  Another interesting possibility is to move away from
the limit cycle regime and investigate the effect of time delay
on the synchronization of coupled chaotic oscillators -  an area
of much current scientific interest \cite{RPK:96}. In principle
this should be possible with our present experimental system
since the basic Chua circuits are capable of displaying chaotic
oscillations. It would be necessary however to employ a faster
delay line. Time delay could facilitate phase synchronization in
view of its significant influence on phenomena like phase slips
\cite{RSJ:00}. A final area of practical interest is the stability
of these limit cycle collective states to external noise. Past
numerical studies indicate that time delay induced higher
frequency states can exhibit metastable behaviour and decay to the
lowest periodic state in the presence of sufficient external noise
\cite{NSK:91}. We do not observe this behaviour since by design
our system has very low intrinsic (thermal) noise in order to
facilitate access to the higher frequency states. It would be
interesting to introduce a suitably designed noise source in the
individual oscillator circuits to experimentally test these
theoretical predictions including the influence of noise on the
threshold for onset of synchronization \cite{Shi:85}.

We acknowledge the important contribution of H.~S. Mazumdar 
in the design, development and construction of the digital delay line.

\end{multicols}


\begin{thebibliography}{10}

\bibitem[*]{GLJadd}
Present Address: EduTron Corp., 5 Cox Rd., Winchester, MA 01890.

\bibitem{KN:87}
Y.~Kuramoto and I.~Nishikawa,
\newblock J. Stat. Phys. {\bf 49}, 569 (1987).

\bibitem{HGEK:97}
A.~Hohl, A.~Gavrielides, T.~Erneux, and V.~Kovanis,
\newblock Phys. Rev. Lett. {\bf 78}, 4745 (1997);
J.~Benford, H.~Sze, W.~Woo, R.~R. Smith, and B.~Harteneck, Phys.
Rev. Lett. {\bf 62}, 969 (1989).

\bibitem{CE:89}
M.~F. Crowley and I.~R. Epstein,
\newblock J. Phys. Chem. {\bf 93}, 2496 (1989);
M.~Yoshimoto, K.~Yoshikawa, and Y.~Mori, Phys. Rev. E {\bf 47},
864 (1993).

\bibitem{CS:93a}
J.~J. Collins and I.~N. Stewart,
\newblock J. Nonlinear Science {\bf 3}, 341 (1993);
M.~A. Branham and M.~D. Greenfield, Nature {\bf 381}, 745 (1996);
M. Kawato and R. Suzuki, J. Theor. Biol. {\bf 86}, 547 (1980); E.
Sismondo, Science {\bf 249}, 55 (1990)

\bibitem{BL:96} A.~A. Brailove and P.S. Linsay, Int. J.
Bifurcation Chaos {\bf 6}, 1211 (1996).

\bibitem{AEK:90}
D.~G. Aronson, G.~B. Ermentrout, and N. Kopell, Physica
(Amsterdam) {\bf 41D}, 403 (1990); P.~C. Matthews and S.~H.
Strogatz, Phys. Rev. Lett. {\bf 65}, (1990); P.~C. Matthews, R.~E.
Mirollo, and S.~H. Strogatz, Physica D {\bf 52}, 293 (1991).

\bibitem{RSJ:98}
D.~V.~Ramana Reddy, A.~Sen, and G.~L. Johnston,
\newblock Phys. Rev. Lett. {\bf 80}, 5109 (1998);
D.~V.~Ramana Reddy, A.~Sen, and G.~L. Johnston, Physica (Amsterdam) {\bf
129D}, 15 (1999).

\bibitem{Strog:98}
S.~H. Strogatz, Nature (London) {\bf 394}, 316 (1998).

\bibitem{HFRPO:00}
R. Herrero, M. Figueras, J. Rius, F. Pi, and G. Orriols,
\newblock Phys. Rev. Lett. {\bf 84}, 5312 (2000).

\bibitem{Chua:93}
M. Komuro, R. Tokunaga, T. Matsumoto, L.O. Chua, and A. Hotta,
\newblock Int. J. Bifurcation and Chaos {\bf 1}, 139 (1991), and
references therein; {\it Chua's Circuit: A Paradigm for Chaos},
edited by R.~N. Madan (World Scientific, Singapore, 1993).

\bibitem{NSK:91}
E.~Niebur, H.~G. Schuster, and D.~Kammen,
\newblock Phys. Rev. Lett. {\bf 67}, 2753 (1991).

\bibitem{KPR:97}
S. Kim, S.~H. Park, and C.~S. Ryu,
\newblock Phys. Rev. Lett. {\bf 79}, 2911 (1997);
M.~K.~S. Yeung and S.~H. Strogatz, Phys. Rev. Lett. {\bf 82}, 648
(1999).

\bibitem{RPK:96} M.~G. Rosenblum, A. S. Pikovsky, and J. Kurths,
\newblock Phys. Rev. Lett. {\bf 76} (1996) 1804.

\bibitem{RSJ:00} D.~V.~Ramana Reddy, A.~Sen, and G.~L. Johnston,
\newblock Physica (Amsterdam) {\bf 144D} (2000) 336.

\bibitem{Shi:85} M. Shiino, Phys. Lett. A {\bf 111}, 396 (1985).

\end{thebibliography}
\end{document}